\documentclass[twocolumn,english,aps,amsmath,amssymb,reprint,nofootinbib,floatfix,showkeys,superscriptaddress]{revtex4-2}

\usepackage[table]{xcolor}
\usepackage{graphicx}
\usepackage{dcolumn}
\usepackage{bm}
\usepackage[colorlinks = true,citecolor=blue,linkcolor = blue,urlcolor=blue]{hyperref}

\usepackage[capitalise]{cleveref}
\usepackage{graphicx}	
\usepackage{ragged2e}
\usepackage{subcaption}
\usepackage{caption}
\usepackage{lineno}
\usepackage{lineno}
\usepackage{amsmath}	
\usepackage{amssymb}	
\usepackage{wasysym}    
\usepackage{mathtools}  
\usepackage{aas_macros}
\usepackage{array,multirow,graphicx}  
\usepackage{multirow}

\renewcommand{\Cref}[1]{\cref{#1}}

\begin{document}

\title{Baryon fraction from the BAO amplitude: a consistent approach to parameterizing perturbation growth}




\author{Andrea Crespi}
\email{a2crespi@uwaterloo.ca}
\affiliation{Waterloo Centre for Astrophysics, University of Waterloo, Waterloo, ON N2L 3G1, Canada}
\affiliation{Department of Physics and Astronomy, University of Waterloo, Waterloo, ON N2L 3G1, Canada}
\author{Will J. Percival}
\affiliation{Waterloo Centre for Astrophysics, University of Waterloo, Waterloo, ON N2L 3G1, Canada}
\affiliation{Department of Physics and Astronomy, University of Waterloo, Waterloo, ON N2L 3G1, Canada}
\affiliation{Perimeter Institute for Theoretical Physics, 31 Caroline St. North, Waterloo, ON NL2 2Y5, Canada \\
The affiliations of the remaining authors are listed in Appendix A.}

\author{Alex Krolewski}
\affiliation{Waterloo Centre for Astrophysics, University of Waterloo, Waterloo, ON N2L 3G1, Canada}
\affiliation{Department of Physics and Astronomy, University of Waterloo, Waterloo, ON N2L 3G1, Canada}

\author{Marco Bonici}%
\affiliation{Waterloo Centre for Astrophysics, University of Waterloo, Waterloo, ON N2L 3G1, Canada}
\affiliation{Department of Physics and Astronomy, University of Waterloo, Waterloo, ON N2L 3G1, Canada}

\author{Hanyu Zhang}
\affiliation{Waterloo Centre for Astrophysics, University of Waterloo, Waterloo, ON N2L 3G1, Canada}
\affiliation{Department of Physics and Astronomy, University of Waterloo, Waterloo, ON N2L 3G1, Canada}

\author{J.~Aguilar\textsuperscript{4}, S.~Ahlen\textsuperscript{5}, A.~Anand\textsuperscript{4}, D.~Bianchi\textsuperscript{6, 7}, D.~Brooks\textsuperscript{8}, E.~Chaussidon\textsuperscript{4}, T.~Claybaugh\textsuperscript{4}, A.~Cuceu\textsuperscript{4}, A.~de la Macorra\textsuperscript{9}, P.~Doel\textsuperscript{8}, S.~Ferraro\textsuperscript{4, 10}, A.~Font-Ribera\textsuperscript{11}, J.~E.~Forero-Romero\textsuperscript{12, 13}, E.~Gaztañaga\textsuperscript{14, 15, 16}, G.~Gutierrez\textsuperscript{17}, J.~Guy\textsuperscript{4}, H.~K.~Herrera-Alcantar\textsuperscript{18, 19}, D.~Huterer\textsuperscript{20, 21}, M.~Ishak\textsuperscript{22}, D.~Joyce\textsuperscript{23}, D.~Kirkby\textsuperscript{24}, T.~Kisner\textsuperscript{4}, A.~Kremin\textsuperscript{4}, O.~Lahav\textsuperscript{8}, C.~Lamman\textsuperscript{25}, M.~Landriau\textsuperscript{4}, L.~Le~Guillou\textsuperscript{26}, M.~E.~Levi\textsuperscript{4}, M.~Manera\textsuperscript{11, 27}, P.~Martini\textsuperscript{25, 28, 29}, A.~Meisner\textsuperscript{23}, R.~Miquel\textsuperscript{11, 30}, S.~Nadathur\textsuperscript{15}, N.~Palanque-Delabrouille\textsuperscript{4, 19}, C.~Poppett\textsuperscript{4, 10, 31}, F.~Prada\textsuperscript{32}, I.~P\'erez-R\`afols\textsuperscript{33}, G.~Rossi\textsuperscript{34}, L.~Samushia\textsuperscript{35, 36, 37}, E.~Sanchez\textsuperscript{38}, D.~Schlegel\textsuperscript{4}, M.~Schubnell\textsuperscript{20, 21}, H.~Seo\textsuperscript{39}, J.~H.~Silber\textsuperscript{4}, D.~Sprayberry\textsuperscript{23}, G.~Tarl\'{e}\textsuperscript{21}, B.~A.~Weaver\textsuperscript{23}, R.~Zhou\textsuperscript{4}, H.~Zou\textsuperscript{40}}



\date{\today}

\begin{abstract}
Galaxy clustering constrains the baryon fraction \(\Omega_\mathrm{b}/\Omega_\mathrm{m}\) through the amplitude of baryon acoustic oscillations and the suppression of perturbations entering the horizon before recombination. This produces a different pre-recombination distribution of baryons and dark matter. After recombination, the gravitational potential responds to both components in proportion to their mass, allowing robust measurement of the baryon fraction (and, with other probes, on the Hubble constant). This constraint is independent of new-physics scenarios altering the recombination background (e.g. Early Dark Energy, EDE). The accuracy of such measurements does, however, depend on how baryons and CDM are modeled in the power spectrum. Previous template-based splittings relied on approximate transfer functions that neglected part of usable information. We present a new method that embeds an extra parameter controlling the balance between baryons and dark matter in the growth terms of the perturbation equations in the \texttt{CAMB} Boltzmann solver. This approach captures the baryonic suppression of CDM prior to recombination, avoids inconsistencies across decoupling, and yields a clean parametrization of the baryon fraction in the linear power spectrum, separating out the simple physics of growth due to the combined matter potential. We implement this framework in a full analysis pipeline using Effective Field Theory of Large-Scale Structure (EFTofLSS) with HOD-informed priors and validate it against noiseless \(\Lambda\)CDM and EDE cosmologies with DESI-like errors. The new scheme achieves comparable precision to previous splittings while reducing systematic biases, providing a more robust route to baryon-fraction measurements. In combination with BBN constraints on the baryon density and Alcock–Paczyński estimates of the matter density, these results strengthen the use of baryon fraction measurements to derive a Hubble constant from energy densities, with future DESI and Euclid data expected to deliver competitive constraints.
\end{abstract}

\maketitle


\section{Introduction \label{sec:intro}}


The flat $\Lambda$CDM model parameters have been constrained to percent level precision by observations of the Cosmic Microwave Background (CMB)~\cite{Planck:2020}, which provides strong evidence for this model. However, tensions with other data have shifted the key questions from those of parameter estimation to consistency tests. One of the strongest tensions is the $\sim4-6\sigma$ tension between local distance‑ladder determinations of the Hubble constant, $H_0 = 73.04\pm1.04\,\mathrm{km\,s^{-1}\,Mpc^{-1}}$, combining Cepheids and supernovae~\cite{Riess-Hubble}, and the CMB‑inferred value, $H_0 = 67.37\pm0.54\,\mathrm{km\,s^{-1}\,Mpc^{-1}}$~\cite{Planck:2020}. Similarly, Baryon Acoustic Oscillation (BAO) measurements yield a comparable low value when combined with the standard ruler of the sound horizon, calibrated using Big Bang Nucleosynthesis (BBN) data~\cite{Cooke2016}. Explanations for the tension range from unidentified systematic errors in the data sets to extensions of $\Lambda$CDM that invoke new physics --- such as Early Dark Energy (EDE)~\cite{Karwal16,Poulin19,Smith20,Hill20} (see also \cite{Lynch24,Mirpoorian25} and \cite{Schoeneberg24} for other early-time modifications that can potentially resolve the Hubble tension). Although the most recent ACT DR6 results~\cite{calabrese2025atacamacosmologytelescopedr6} alone seem to disfavor the EDE scenario, when combined with supernova data and DESI DR2 BAO~\cite{poulin2025impactactdr6desi}, the hint of a non-zero EDE disappears. Therefore, independent and complementary measurements of $H_0$ are essential to understand the origin of this discrepancy.

Recently, several methods have been developed to determine \(H_0\) without relying on calibrating distance ladders with the sound horizon at the drag epoch. Philcox et al.~\cite{Philcox22} (using methods developed in \cite{BaxterSherwin21,Farren22,Philcox21c}) infered \(H_0\) from the matter–radiation equality scale using BOSS full-shape power spectra and CMB lensing, marginalizing over the sound horizon to remove acoustic information. Bahr-Kalus et al.~\cite{bahrkalus2025} directly detect the turnover in the matter power spectrum and use it as a standard ruler, yielding a sound-horizon–free estimate of the Hubble constant. Zaborowski et al.~\cite{Zaborowski_2025} extend this strategy by analyzing the full-shape clustering of multiple tracers, rescaling the power spectrum to explicitly marginalize over the sound horizon, and combining with CMB lensing and uncalibrated SNe to reach sub-3\% precision. In all cases, the calibration depends on equality physics or direct shape information rather than on the sound horizon at drag epoch, making the resulting \(H_0\) measurements independent of recombination-era acoustic physics.

In a recent paper~\cite{krolewski_H0}, Krolewski et al. presented a new method for inferring the Hubble constant that avoids using CMB power-spectrum data and does not require calibrating the BAO with a predetermined sound-horizon scale. This method combines a purely geometric measurement of $\Omega_m$ derived from the Alcock–Paczyński effect applied to BAO (or to cosmic voids~\cite{Lavaux_2012}), together with the cosmological value of $\Omega_bh^2$, constrained by BBN data. The missing piece to determine $H_0$ is an independent measurement of the baryon fraction $f_b = \Omega_b/\Omega_m$. In their companion paper~\cite{krolewski_fb} (hereafter KP25), Krolewski \& Percival introduced a new method to infer the baryon fraction from the amplitude of the baryonic signal in the galaxy power spectrum. This baryonic signature consists of two effects: the acoustic oscillations imprinted in the range $k \sim 0.02-0.3 \; h \,\mathrm{Mpc}^{-1} $ and a scale‑dependent suppression on sub-sound‑horizon scales. In particular, Fig.~1 in KP25 shows that the BAO amplitude can recover the true baryon fraction $f_b$, independent of whether the underlying cosmology follows $\Lambda$CDM or an EDE model. Thus, it provides the key ingredient for a energy density based measurement of $H_0$.

In this work, we develop a more robust method for measuring the baryon fraction than that adopted by KP25, by introducing an additional free parameter to separate the growth contributions of baryons and cold dark matter in the perturbation equations. We begin by contextualizing the problem in Section~\ref{sec:fb}, present the new method in Section~\ref{sec:new_method}, and describe how it can be integrated into a pipeline for measuring $H_{0}$ in Section~\ref{sec:pipeline}. In Section~\ref{sec:validation-tests}, we provide a detailed comparison between the original KP25 template method and our perturbation-level implementation on noiseless theory vectors spanning $\Lambda$CDM and EDE cosmologies, and our conclusions are summarized in Section~\ref{sec:conclusions}. The application of this method to the Dark Energy Spectroscopic Instrument (DESI; \cite{DESI2016}), Data Release 1 (DR1; \cite{desicollaboration2025datarelease1dark}) sample is provided in our companion paper Krolewski et al. (in prep).

\section{Measuring the baryon fraction from growth} \label{sec:fb}

Within galaxy redshift surveys, the imprint left by baryons is one of the most robust physical signatures that can be extracted. Classical BAO analyses (e.g. \cite{Alam-eBOSS:2021}) have focused on measuring the observed positions of the feature along and across the line of sight, which scale as $D_H/r_\mathrm{d}$ and $D_M/r_\mathrm{d}$, with $D_H(z)\equiv c/H(z)$, $D_M(z)\equiv(1+z)D_A(z)$, and $r_\mathrm{d}$ the comoving sound horizon at the drag epoch. The sound horizon depends non-trivially on $\Omega_b$, $\Omega_m$, and $h$, requiring a complex physical modeling. Nevertheless, it is predicted to very high precision once the cosmology is fixed.
In linear theory, the cosmic evolution of density perturbations is encapsulated in scale-dependent transfer functions: each component --- baryons, cold dark matter, photons, neutrinos --- has its own $T_i(k)$ that translates primordial fluctuations into present-day amplitudes. For an energy-density dominated by matter, the overall transfer function is 
\begin{equation}
T(k) = \frac{\Omega_b}{\Omega_{bc}}\, T_b(k) + \frac{\Omega_c}{\Omega_{bc}}\, T_c(k),
\label{tot_transfer_definition}
\end{equation}
where \(\Omega_{bc} \equiv \Omega_b + \Omega_c\),
with the baryonic and cold-dark-matter components entering in proportion to their energy-density fractions. This function, feeds directly into the linear clustering power spectrum:
\begin{equation}
P(k, z) = A_s\, \left(\frac{kh}{0.05}\right)^{n_s -1} \,2\pi^2k\,h^4\,T^2(k)\, D^2(z) \; .
\end{equation}
Eq.~\ref{tot_transfer_definition} exhibits an explicit and simple dependence on the fraction $\Omega_b/\Omega_{bc}$. 

To extract just this signal, we can consider replacing the weighting between \(T_b\) and \(T_c\) with a free parameter (hereafter $\gamma_b$), such that the weights for the two terms are $\gamma_b$ and $1-\gamma_b$. 
$\gamma_b$ is an estimator of the ratio between baryon and baryon+CDM densities coming from just the growth of perturbations.

Before photon decoupling the two transfer functions, \(T_c\) and \(T_b\), differ strongly, with \(T_c\) being smooth as a function of \(k\), while in the limit \(k \,r_\mathrm{d} \gg 1\), the baryonic term can be approximated by 
\begin{equation}
    T_b(k)=\alpha_{b}\frac{\sin(k r_\mathrm{d})}{k r_\mathrm{d}}\,\mathcal{D}(k)\;.
\end{equation}
\(\mathcal{D}(k)\) represents the Silk damping effect, the photon diffusion effect that removes energy from baryon overdensities, while $\alpha_b$ accounts for the suppression produced by the declining sound speed and the reduced growth between matter–radiation equality and the drag epoch. $\alpha_b$ is governed mainly by $f_b$ with only a mild residual dependence on $\Omega_m h^{2}$.  After the drag epoch, the baryonic and CDM perturbations grow together under the summed gravitational potential, leaving the shape of the total transfer function fixed and altering only its overall amplitude, which is scale-independent in the linear regime. Since $f_b$ is constant with redshift in standard cosmology, $T_b(k)$ and $T_c(k)$ can be evaluated at any convenient epoch (rescaling their amplitudes for the correct redshift), and recombined as in Eq.~\ref{tot_transfer_definition}. This freedom can be used to potentially maximize the contrast between the two transfer functions and improve the precision of the $\gamma_b$ measurement that tried to decompose the matter power spectrum into the two components.

As explained in KP25, the baryon and CDM transfer functions used in Eq.~\ref{tot_transfer_definition} must be chosen with care. Taking the \(T_b\) and \(T_c\) outputs from a Boltzmann solver, such as \texttt{CAMB}~\cite{Lewis_2000} or \texttt{CLASS}~\cite{lesgourgues2011cosmiclinearanisotropysolving}, at redshift \(z=0\) would be completely ineffective. You would see both a suppression of BAO signal in \(T_b\) due to the growth of structures, and the presence of oscillations in \(T_c\) due to the coupled evolution of CDM and baryon in their combined gravitational potential. I.e.\ the two transfer functions are very similar at low redshift, removing any constraining power on the parameter \(\gamma_b\). However, it is also not a good choice to take the transfer functions at high redshift when they are maximally different, namely at drag epoch redshift \(z_\mathrm{d}\sim1060\). The acoustic oscillations do not stop traveling immediately at \(z_\mathrm{d}\); instead the residual velocity at decoupling allows them to propagate, with a decreasing speed, until \(z\sim600\). Unfortunately, at that redshift CDM already starts to clump around the baryons, removing part of the oscillatory signal from \(T_b\), making the transfer functions at \(z\sim600\) also not optimal. To avoid these problems, KP25 opted to use the Eisenstein and Hu fitting formula \cite{Eisenstein-Hu} for \(T_c\), the smooth CDM component, while they defined the baryonic \(T_b\) as the density-weighted difference between \texttt{CLASS} baryon+CDM transfer function at \(z=0\) and Eisenstein and Hu (hereafter EH) CDM transfer function, according to
\begin{equation}
T_b(k) = \left( T_{cb}(k, z=0) - \frac{\Omega_c}{\Omega_{cb}}T_c^\mathrm{EH}(k) \right) \frac{\Omega_{cb}}{\Omega_b} \; .
\end{equation}
This combines the maximal split of the signal between transfer functions with the final positions of the acoustic waves. However, as we are going to further explore in Sec.~\ref{sec:new_method}, this splitting choice has two major limitations. First, the numerical precision of EH transfer functions --- within \(3.6\%\) for Planck2018 fiducial cosmology --- puts a bound on the precision of \(\gamma_b\) measurements.  Second, this approach fixes the shape of \(T_c\) as it is at the drag epoch; then it evaluates the effect of different \(\gamma_b\) on clustering simply recombining this \(T_c\) with the baryon components rescaled for the different baryon fraction. It neglects the baryonic suppression of CDM perturbations that occurs prior to photon–baryon decoupling, which changes $T_c$. This omission reduces physical self‑consistency and discards usable information on $\gamma_b$ from the shape of the power spectrum.

For this analysis, we assumed a single set of bias parameters when comparing the galaxy field with both the CDM and baryon density fields. However, the differential velocity and density of the baryonic field relative to the CDM field at early times can induce bias contributions, effectively introducing two additional terms, \(b_{\delta_{bc}}\) and \(b_{v^2_{bc}}\). These terms depend on the physics of galaxy formation and are scale-dependent. Similarly, compensated isocurvature perturbations (CIPs), which are primordial fluctuations in the baryon density compensated by corresponding fluctuations in the CDM density to preserve an unperturbed total matter density, can also generate scale-dependent bias term. Although these effects can, in principle, be modeled, they are currently negligible given the precision of existing data. A detailed treatment of these contributions will be considered in future extensions of this work. A qualitative discussion of the impact of relative velocity effects can be found in Appendix A of the companion paper \cite{krolewski25_H0}.

\section{Modeling the growth through perturbation evolution}
\label{sec:new_method}

We propose a new approach that embeds $\gamma_b$ directly in the linear perturbation equations of \texttt{CAMB}. The baryon and CDM densities are rescaled according to the new parameter, so the full temporal evolution of both fluids responds coherently to the chosen $\gamma_b$. The thermal history and the adiabatic initial conditions, governed by the background cosmological parameters, are unaltered. This means that the new parameter is able to extract the dependence of perturbation growth only on the baryon fraction. 
This type of approach is not new. Several works in the literature have split parameters into two distinct counterparts, each responsible for different physical phenomena, motivated by different goals. For example, the separation between a total matter energy density parameter for growth and one for geometry to test the \(\Lambda\)CDM model in~\cite{Andrade_2021}, or the splitting of the baryon density into a parameter controlling the ionization history and another governing the rest of the baryon effects in~\cite{Chu_2005}.
There are different ways to separate the dependence on the growth in the physical processes and in their equations, each encapsulating distinct physical aspects of the evolution of perturbations. We consider a number of options to better explain why our chosen method is the most coherent and complete after first reviewing the physical equations governing the evolutions of perturbations.

\subsection{Evolution of perturbations}
\label{Sec:Evolution of perturbations}

\texttt{CAMB}\footnote{Particularly valuable for understanding the equations for the evolution of the perturbations and their numerical implementation were Prof. Antony M. Lewis’s PhD thesis, \textit{Geometric Algebra and Covariant Methods in Physics and Cosmology} \cite{AntonyLewisThesis}, as well as his \textit{CAMB Notes} \cite{camb_notes}.} numerically integrates the linearized Einstein–Boltzmann system in synchronous gauge, evolving all species in conformal time \(\tau\) so that the gravitational potentials sourced by one component instantaneously feed back on every other.  
The background expansion obeys the Friedmann equation
\begin{equation}
\mathcal{H}^2 \equiv \left(\frac{a'}{a}\right)^2 = \frac{8\pi G}{3}\,a^2\sum_i \bar\rho_i(\tau)\,,
\label{eq:Friedmann}
\end{equation}
where a prime denotes $\mathrm{d}/\mathrm{d}\tau$ and the sum runs over photons, massless and massive neutrinos, cold dark matter, baryons, and dark energy.  
Metric perturbations are described in synchronous‐gauge by the two scalar potentials $h$ and $\eta$.  The Hamiltonian and momentum equations result in
\begin{equation}\begin{split}
k^2\eta - \tfrac12\mathcal{H}h' &= 4\pi Ga^2\sum_i\bar\rho_i\delta_i \\
k\eta' &= 4\pi Ga^2\sum_i\left(\bar\rho_i+\bar P_i\right)v_i\,,
\label{eq:eta_prime}
\end{split}\end{equation}
with density contrast $\delta_i\equiv\delta\rho_i/\bar\rho_i$ and pressure $P_i$.  
Cold dark matter evolves according to
\begin{equation}
\delta_c'=-\tfrac12h', \qquad
v_c'=0\,,                \label{eq:cdm}
\end{equation}
while baryons are tightly coupled to photons through Thomson scattering before decoupling. This tight coupling results in a high opacity \( \tau_c^{-1}\), and in this regime, the full Boltzmann hierarchy for photons becomes stiff and extremely inefficient. To address this, \texttt{CAMB} implements a tight-coupling approximation by expanding in \( \tau_c \), valid when \( \epsilon \equiv \max(k\tau_c, \mathcal{H}\tau_c) \ll 1 \):
\begin{equation}\begin{split}
\delta_b' &=-\tfrac12h'-kv_b \\
v_b' &= \frac{1}{1+R} \Big[ kc_s^2 \delta_b + \frac{k}{4} R(\delta_\gamma - 2\beta_2 \pi_\gamma) - \mathcal{H} v_b \\
&\quad + \left( \frac{2\mathcal{H}}{1+R} + (\ln \tau_c^{-1})' \right) \frac{3R}{4(1+R)} \Delta \\
&\quad - \frac{R\tau_c}{4(1+R)^2} \Big( 4(\mathcal{H}' + \mathcal{H}^2)v_b  \\ &+k(2\mathcal{H} \delta_\gamma + \delta_\gamma' - 4c_s^2 \delta_b') \Big) \Big]\,. \label{eq:baryon_euler}
\end{split}\end{equation}
Here $\tau_c$ is the optical depth, $R\equiv 4\bar\rho_\gamma/3\bar\rho_b$, and $c_s$ is the baryon sound speed. 
The leading term \( kc_s^2 \delta_b \) represents the baryon pressure term, while \( \frac{R}{4}(\delta_\gamma - 2\beta_2 \pi_\gamma) \) encodes the momentum input from photon overdensities \( \delta_\gamma \) and the anisotropic stress \( \pi_\gamma \), which contributes to quadrupole radiation pressure. The coefficient \( \beta_2 \) accounts for the polarization correction to the quadrupole moment (\( \beta_2=1 \) in the absence of polarization). The term \( -\mathcal{H} v_b \) reflects Hubble drag due to the expansion of the Universe. The \( \Delta \) term captures the photon-baryon slip, i.e., the relative velocity between photons and baryons. \( \pi_\gamma \) quantifies the anisotropic stress in the photon distribution and is computed from the baryon velocity and the metric shear \( \sigma = (h' + 6\eta_s')/(2k) \). These approximations are necessary to ensure numerical stability and continuity at early times, particularly for small scales where the photon-baryon fluid exhibits acoustic oscillations.

They transition, then, to a epoch of non-tight coupling where
\begin{equation}
v_b' \approx -\mathcal{H} v_b + c_s^2 k \delta_b + \frac{k}{4}R \delta_\gamma \,,
\end{equation}
which remains valid even after photon decoupling to account for residual baryon-photon interactions from drag epoch onwards.

Photon evolution is described by the temperature multipoles $\Theta_\ell$, which satisfy the Boltzmann hierarchy with a baryon-photon scattering term controlled by \(\tau_c'\).
Massless neutrinos have an analogous system of equations for the multipoles $F_\ell$, but without the scattering term $\tau_c'$.  
Massive neutrinos are integrated in momentum shells with a similar hierarchy weighted by the Fermi–Dirac distribution.  
Anisotropic stress from free‐streaming species feeds back on the metric through Eq.~\ref{eq:eta_prime}, damping potential decay inside the horizon.

Evolving this set of equations all together is essential because each species continuously sources the metric potentials that, in turn, feed back on every other species. Photon–baryon pressure alters the wells into which cold dark matter falls; CDM gravity reshapes the acoustic oscillations of the photon–baryon fluid; free‑streaming neutrinos and dark energy likewise modify the same potentials at later times. This tight interdependence means that there is no perfect, redshift‑zero transfer function splitting: the baryonic imprint accumulates incrementally at every integration step, inseparable from the concurrent CDM response. Quantifying their individual contributions must therefore compare full, self‑consistent integrations rather than isolating them at the end.

\subsection{Modification Strategy}

We introduce a new parameter, the growth baryon fraction \(\gamma_b\), allowing this to be different from the cosmological baryon fraction $f_b$ fixed by the background parameters. This allows the growth of perturbations to respond to a different balance of baryonic and dark-matter contributions than is actually present. This is clearly unphysical, but is only used as a mechanism to split the information from the different physical processes. In a fit to data it allows us to extract only the information on the baryon fraction from perturbation growth by considering only the constraint on \(\gamma_b\). 
To implement this strategy, we introduce \(\gamma_b\) only in evolutionary equations of perturbations that depend on the growth. However, there are still choices to be made about exactly how the modification is implemented.

One option would be to introduce \(\gamma_b\) only in the evolutionary equations of perturbations after decoupling. Because most of the structure formation occurs after recombination, this approach is the natural analogue of the original splitting between transfer functions considered in Eq.~\ref{tot_transfer_definition}, while still allowing for the fully coupled evolution of the various components. To implement this, each of the evolution equations previously introduced was altered at a decoupling time $t_{\mathrm{dec}}$: for $t < t_{\mathrm{dec}}$ they contain only the cosmological baryon fraction, whereas for $t > t_{\mathrm{dec}}$ the cosmological value is replaced by growth $\gamma_b$, but only where it contributes to the growth of structure. Obviously, this procedure introduces a degree of inconsistency because the evolution equations are altered instantaneously.
Although it does decompose the primary physical processes, we find a number of undesirable features with this approach, suggesting that the information has not been smoothly separated. In the extreme case of a null growth baryon fraction the final matter power spectrum retains residual baryonic oscillations generated prior to decoupling. Conversely, for very large growth baryon fractions, the power spectrum exhibits a higher overall amplitude than one would expect because it also preserves the imprint of the baryonic suppression experienced by CDM perturbations before decoupling. These behaviors arise from the fact that part of the information on the baryon/CDM contribution to structure growth originates in the pre-recombination epoch. With this approach, the contributions are well separated at decoupling, but the components that were not separated prior to decoupling re-contaminate each other afterward due to the coupled evolution of all the equations at any cosmic time. As a result, after decoupling, the evolution of the perturbations show a phase that lacks a clear physical interpretation, leading to the undesired situations described above.

A second strategy would be to allow the metric perturbations to depend on the growth baryon fraction at all cosmic times, while all other equations are split at decoupling, depending on the cosmological baryon fraction before decoupling and the growth baryon fraction afterwards. In this setup, no change in the cosmological baryon‑fraction information occurs at decoupling for the joint baryon and CDM perturbations. This approach, however, is even less self‑consistent than the first: not only are the evolution equations of the various species switched instantaneously, but one component remains permanently modified, experiencing no change across the decoupling threshold.

The reason for considering the above scenarios is that we want to ensure the growth baryon fraction used in the equations does not affect the physics before recombination. However, what they show is that an abrupt transition leaves undesirable signatures in the clustering pattern. 

\subsection{The new approach}

\begin{figure*}[!htp]
    \centering
    \begin{subfigure}{0.5\textwidth}
        \includegraphics[width=\linewidth]{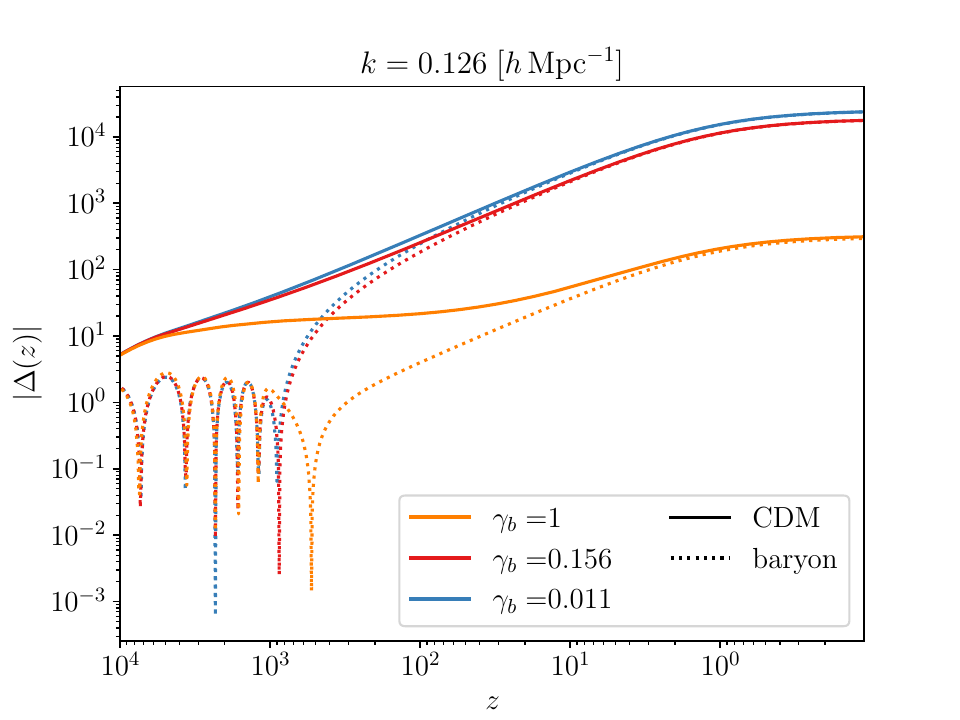}
    \end{subfigure}
    \hspace{-7mm}
    \begin{subfigure}{0.5\textwidth}
        \includegraphics[width=\linewidth]{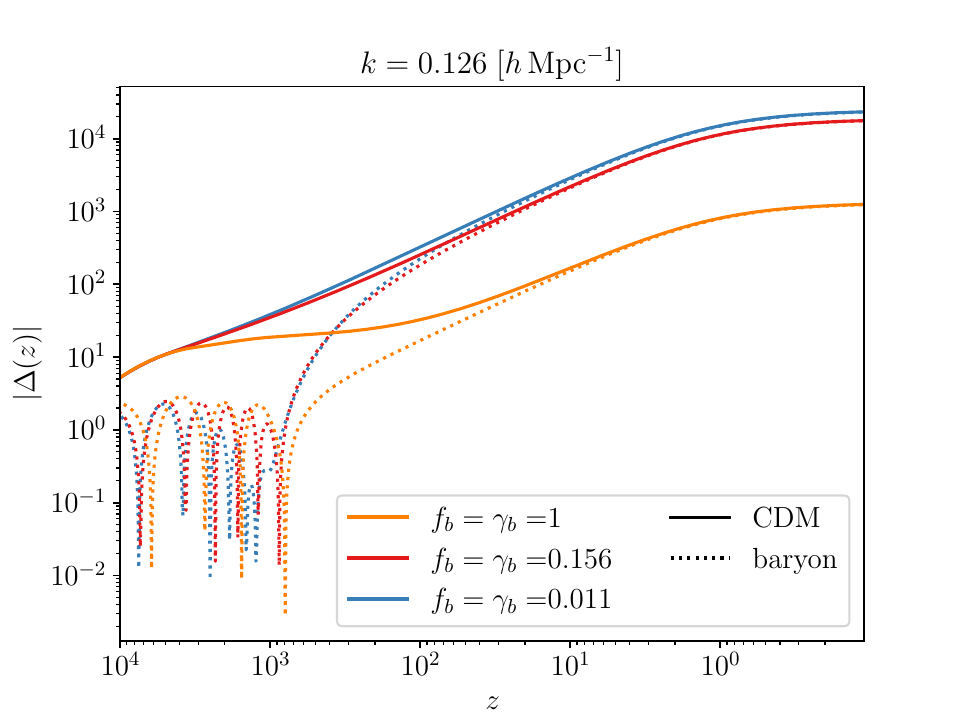}
    \end{subfigure}
    \vspace{-2mm}
    \caption{\justifying LEFT Redshift evolution of the absolute density contrasts \(|\Delta(z)|\) for baryons (dotted lines) and CDM (solid lines) at fixed comoving scale \(k = 0.126\,h\,\mathrm{Mpc}^{-1}\).  
    Three growth-baryon-fraction values are shown: \(\gamma_b\approx0\), \(\gamma_b=0.156\) (fiducial cosmological value), and \(\gamma_b=1\).  
    All lines share identical adiabatic initial conditions and the recombination redshift \(z\simeq1100\). Before decoupling the baryon perturbations undergo acoustic oscillations that are mostly independent of \(\gamma_b\). After decoupling the growth rate diverges: when $\gamma_b \approx 1$, perturbations evolve solely through baryons, showing delayed growth; when $\gamma_b \approx 0$, perturbations evolve solely through CDM, which grows slightly faster and baryons collapse immediately after recombination.
    RIGHT Same as on the left, but changing the global baryon fraction $f_b$ as well as the growth-baryon fraction $\gamma_b$.
    The figure shows the scenario where the global baryon fraction is varied for three values: $f_b=\gamma_b\approx0$, $f_b=\gamma_b=0.156$ (fiducial), and $f_b=\gamma_b=1$. 
    While all lines share identical adiabatic initial conditions, they have distinct thermal and growth histories: changing the global baryon fraction not only alters the relative clustering of baryons and CDM, but also affects the recombination epoch and the subsequent evolution of perturbations, leading to significantly different BAO patterns. Since we wish to extract the growth information only while leaving the position of the peaks unchanged, we use the new parameter $\gamma_b$ to constrain the baryon fraction rather than $f_b$.}
    \label{fig:treu_vs_wrong_delta_z_evol}
\end{figure*}

We favour a scenario where we replace the cosmological baryon fraction with its growth counterpart in the equations at all cosmic times, including the pre‑recombination era. Importantly, we implemented this by carefully considering all the relevant equations in \texttt{CAMB} so that the growth baryon fraction modifies only structure formation; the thermal history of the Universe remains entirely set by the standard cosmological parameters. This approach removes any discontinuity in the dynamics of perturbations and is therefore self‑consistent, including baryon induced suppression of CDM perturbations before decoupling.
The collapse of matter overdensities, and the subsequent growth of structure, is a delicate interplay between gravitational collapse, expansion of the Universe and mutual interaction between different species. Hence, we have to modify most of the terms in the equations of Sec.~\ref{Sec:Evolution of perturbations}. First of all, different baryon and CDM abundances would alter the expansion of the Universe through the Friedmann equation, Eq.~\ref{eq:Friedmann}. Rescaling the baryon and CDM densities with \(\gamma_b\), but preserving the total matter density, we obtain:
\begin{equation}\begin{split}
\mathcal{H}^2 =\frac{8\pi G}{3}\,a^2\left( \frac{\gamma_b}{f_b}\bar\rho_b(\tau) + \frac{1-\gamma_b}{1-f_b}\bar\rho_c(\tau) +
\sum_i \bar\rho_i(\tau)\right) \,,
\end{split}\label{eq:mod_Friedmann}
\end{equation}
where the final sum runs over the remaining species. The response of the metric perturbation is also altered in Eq.~\ref{eq:eta_prime} to 
\begin{equation}\begin{split}
k^2\eta - \tfrac12\mathcal{H}h' &= 4\pi Ga^2\left( \frac{\gamma_b}{f_b}\bar\rho_\mathrm{b}\delta_b+\frac{1-\gamma_b}{1-f_b}\bar\rho_\mathrm{c}\delta_c+\sum_i\bar\rho_i\delta_i \right)\\
k\eta' &= 4\pi Ga^2\left(\frac{\gamma_b}{f_b}\bar\rho_\mathrm{b}v_b+\sum_i\left(\bar\rho_i+\bar P_i\right)v_i\right)\,,
\label{eq:mod_eta_prime}
\end{split}\end{equation}
where, in the second line, there are no CDM terms since in the synchronous gauge it has zero velocity. The equations for \(\delta'_c\) and \(\delta'_b\) are not modified further since \(h'\) already accounts for the correct growth baryon fraction. Eq.~\ref{eq:baryon_euler} for the baryon velocity needs to be treated carefully. Those terms coming from the thermal history of the Universe are left unchanged. This means that the terms responsible for the position of the acoustic peaks do not change, whereas those giving the suppression of baryon growth are modified according to \(\gamma_b\). Parameters \(R\), \(c_s\) and \(\tau_c\) are not modified, and are fully determined by the background cosmology. All other terms in the equations for \(\mathcal{H}\), \(\delta_b\) and \(\pi_\gamma\) are modified since the evolution of these terms is modified with the growth baryon fraction.
With these modifications to the equations, we obtain the evolution of perturbations as shown in the first panel of Fig.~\ref{fig:treu_vs_wrong_delta_z_evol}. The initial amplitudes of both baryon and CDM perturbations are identical for any value of $\gamma_b$, and recombination still occurs at $z \simeq 1100$, as expected given that the growth baryon fraction is not used in computing the initial conditions or the thermal history. The evolution of the perturbation amplitudes, however, changes significantly with $\gamma_b$. For baryons, prior to decoupling their oscillatory behaviour is set by their tight coupling to photons; because the thermal history is left unchanged with respect to the background cosmology, the baryon sound speed is unaffected, and the pre‑decoupling evolution of baryonic perturbations is therefore essentially independent of $\gamma_b$. Instead, after decoupling, they behave quite differently.

To help to visualize this behaviour we consider two extreme scenarios.

When \(\gamma_b\approx1\), the evolution of perturbations is driven solely by the baryonic component, enhanced to match the entire matter budget.
Since before recombination baryonic perturbations are pressure-supported, CDM perturbations do not grow prior to recombination. After decoupling the baryons do not respond to the deep CDM gravitational potential wells present in the fiducial case, and thus we observe in the baryonic perturbations a significant delay before they begin to grow. Only once they start growing, the CDM perturbations also begin to collapse, reaching the same amplitude as the baryonic perturbations. 

In the opposite scenario, when \(\gamma_b\approx0\), the evolution of perturbations is determined solely by the CDM component.
As a result, CDM perturbations grow slightly faster, since they do not respond to the baryonic suppression present before decoupling in the fiducial case. Baryonic perturbations, which remain pressure-supported prior to recombination, collapse immediately afterward
into the CDM potential wells.

We compare this procedure to that of modifying a single universal baryon fraction in the second panel of Fig.~\ref{fig:treu_vs_wrong_delta_z_evol}. We clearly see the effect of the different thermal histories with distinct baryon signature at $z \lesssim 1100$ leading to differences in the positions of the BAO peaks. Since these modifications include baryonic effects on the thermal history of the Universe and not only on structure growth, they fall outside the scope of this work and are therefore discarded.

\begin{figure}[!ht]
    \centering
    \includegraphics[scale = 0.55]{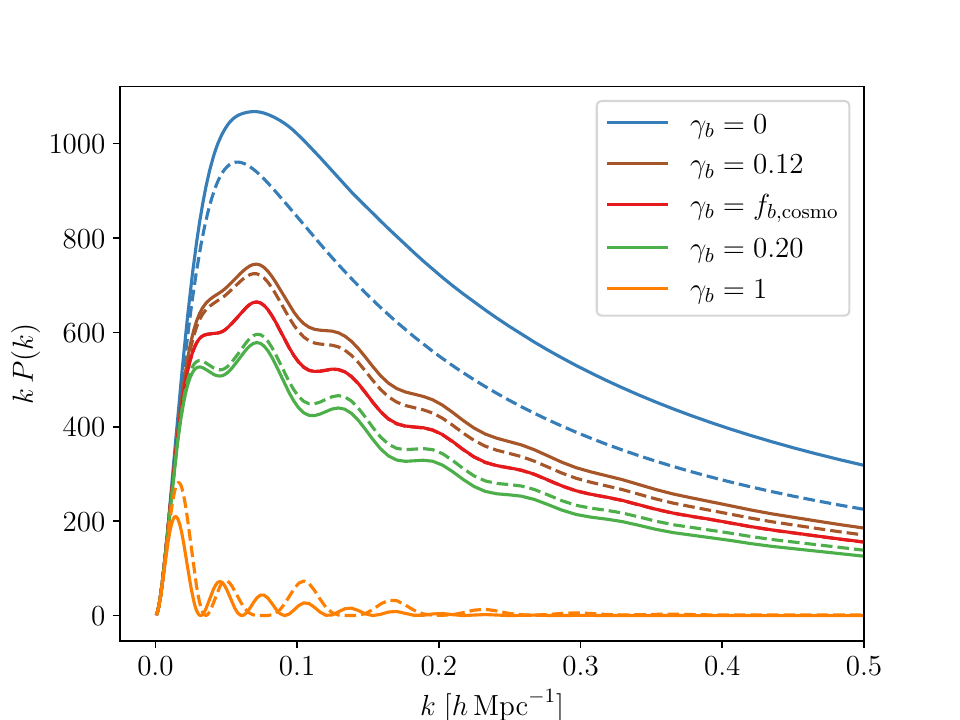}
    \caption{\justifying Comparison between the linear power spectra obtained with our new method (solid lines) and the KP25 splitting scheme (dashed lines), for different values of \(\gamma_b\) ranging from 0 to 1, including the fiducial cosmological value of 0.156. The two methods encapsulate different physical effects in the definition of the \(\gamma_b\) parameter, leading to visible differences in amplitude, especially at the extremes. At \(\gamma_b=0\), our method removes the baryon-induced suppression on the CDM transfer function prior to the drag epoch. At \(\gamma_b=1\), it also eliminates the CDM-driven growth in baryon oscillations during drag epoch that is retained in the EH transfer functions used by KP25.}
    \label{fig:old_vs_new_pklin}
\end{figure}

Fig.~\ref{fig:old_vs_new_pklin} shows a comparison between the linear power spectrum obtained with our new approach (solid lines) and that obtained using the previous KP25 splitting scheme (dashed lines), for a range of \(\gamma_b\) values between 0 and 1. 
For values close to the cosmological baryon fraction, the differences are relatively small, but they become increasingly significant as we move toward either extreme. For \(\gamma_b=0\), the large difference in amplitude comes from the suppression of CDM perturbation growth caused by baryons prior to the drag epoch. 
When \(\gamma_b=1\), we observe a small difference in the amplitude of the acoustic peaks due to residual CDM-induced growth in the transfer function approach that is not present if we model the different growth at source.

\section{Analysis pipeline}
\label{sec:pipeline}
Having established a consistent framework to implement the growth baryon fraction \(\gamma_b\) within the perturbation evolution equations, we now turn to its application in the full inference framework of the energy density based \(H_0\) measurement~\cite{krolewski_H0}. In the following, we describe the structure of the analysis, the role of nuisance and cosmological parameters, and the use of HOD-informed priors to mitigate degeneracies and projection effects.
\subsection{Fitting procedure: full--shape + compressed post-reconstruction BAO}
\label{sec:fitting-procedure}

As our baseline analysis, we simultaneously fit $\gamma_b$ to the power spectrum multipoles,
and also use geometric information from the post-reconstruction BAO scaling parameters.
Since the information encoded in the shape of the power spectrum and that from BAO is not statistically independent, the analysis must be performed allowing for the covariance between the two---using the standard EZmock-based covariance matrix between the pre-reconstruction power spectrum multipoles and the compressed BAO parameters $\alpha_\parallel$ and $\alpha_\perp$.

The amplitude of the BAO in the post-reconstruction data also depends on \(\gamma_b\).
In our companion paper~\cite{krolewski25_H0}, we tested the impact of incorporating the post-reconstruction measurement of \(\gamma_b\) into our pipeline by adding an additional data point and recomputing the covariance accordingly. However, there is no significant improvement in the error on  \(\gamma_b\), since the \(\gamma_b\) constraint from our pipeline and that obtained from post-reconstruction BAO for the adopted simulations are correlated at the \(\sim60\%\) level. Consistently, the gain in constraining power is only about \(\sim6\%\). We therefore opted not to include this measurement in our baseline pipeline to avoid introducing additional complexity, particularly regarding the impact of different reconstruction techniques on \(\gamma_b\).

\begin{figure*}[!htp]
    \centering
    \begin{subfigure}{0.5\textwidth}
        \includegraphics[width=\linewidth]{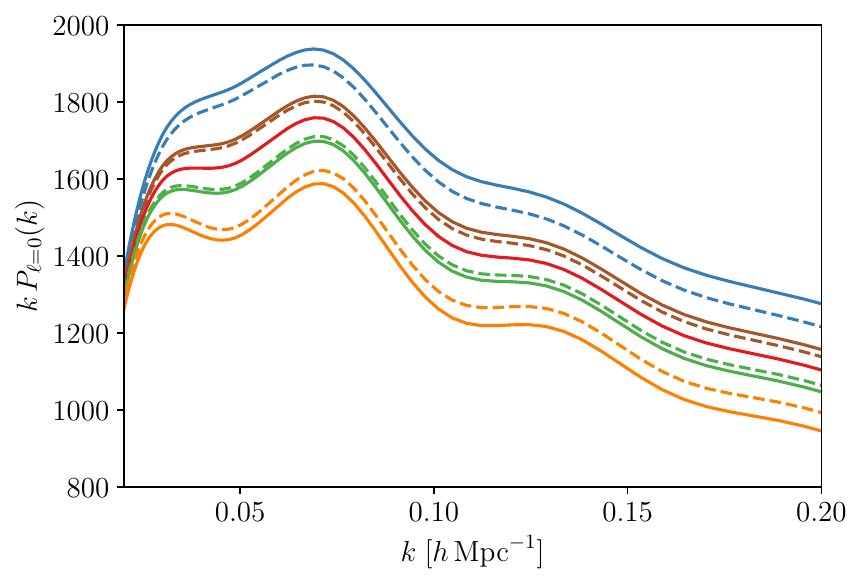}
    \end{subfigure}
    \hspace{-3.5mm}
    \begin{subfigure}{0.5\textwidth}
        \includegraphics[width=\linewidth]{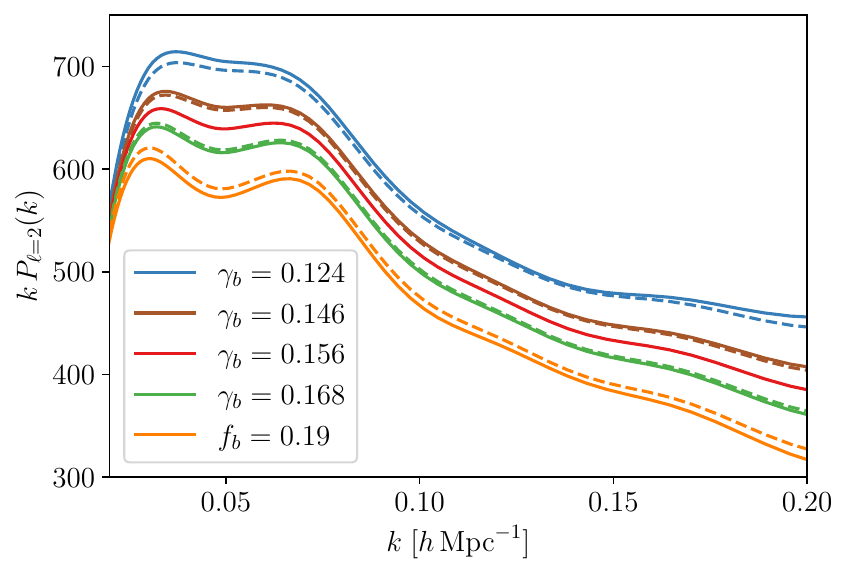}
    \end{subfigure}
    \vspace{-2mm}
    \caption{\justifying Comparison of the impact of different values of \(\gamma_b\) on the monopole and quadrupole of the power spectrum computed using EPT for both the KP25 splitting (dashed lines) and our new method (solid lines). The different values of \(\gamma_b\) correspond to \(\pm1\sigma\) and \(\pm3\,\sigma\) around the fiducial value of 0.156.}
    \label{fig:Pk_ells_vs_gamma_b}
\end{figure*}

As a result, our analysis includes an additional parameter beyond the standard set of cosmological and nuisance parameters adopted in the Effective Field Theory of Large-Scale Structure framework~\cite{Baumann12,Carrasco12}. \(\gamma_b\) exclusively modifies the linear power spectrum, which is then used as input of perturbation theory pipeline implemented in \texttt{velocileptors}~\cite{Chen_2020,Chen_2021}. The code computes, in both Lagrangian (LPT) and Eulerian (EPT) formalisms, the real- and redshift-space power spectra and correlation functions of biased tracers using 1-loop perturbation theory (including effective field theory counterterms and up to cubic-order biasing).
In Fig.~\ref{fig:Pk_ells_vs_gamma_b} we show the effect of \(\gamma_b\) on the monopole and quadrupole of the power spectrum, computed with EPT and a fiducial set of EFT parameters, for both our new method and the old KP25 splitting. 
The variation in \(\gamma_b\) is smaller than in the linear power spectrum and corresponds to \(\pm1\sigma\) and \(\pm3\sigma\) around the fiducial value. We do not show the effect for more extreme values, since the power-spectrum amplitude would enter a regime where the EFT approximation is no longer valid, leading to results that are not reliable and difficult to interpret physically.

The perturbative model introduces a full set of nuisance parameters to account for nonlinear and bias-related effects:
\begin{itemize}
    \item Bias terms, describing the relation between the galaxy distribution and the underlying matter density field. It includes linear bias \(b_1\), quadratic bias \(b_2\), tidal bias \(b_s\), and a third-order bias term \(b_3\).
    \item Counterterms, arising from the effective field theory approach itself, which compensate for the counter effect from the small-scale modes in the perturbative expansion. These are of the form \(\alpha_i k^2P_\mathrm{lin}(k)\), with \(\alpha_i\) treated as free parameters, usually marginalized over.
    \item Stochastic terms, which model shot noise and other contributions such as the Finger-of-God dispersion. These are generally constant or weakly scale-dependent, and may include corrections proportional to \(k^2\).
\end{itemize}
An alternative parametrization of the nuisance parameters is also available in \texttt{velocileptors}, offering a more physically intuitive approach to prior specification. In the default configuration, the galaxy bias parameters are scaled with \(\sigma_8(z)\) as $(1 + b_1)\sigma_8(z)$, $b_2\sigma_8(z)^2$, $b_s\sigma_8(z)^2$, and $b_3\sigma_8(z)^3$. We will refer to this choice as the physical basis.
As discussed in~\cite{maus2024analysisparametercompressionfullmodeling}, this parameter basis more effectively captures the dependence of the monopole on the bias parameters, better aligning with the observables given the high degeneracy between \(\sigma_8\) and the biases. Other works, such as 
\cite{maus2024comparisoneffectivefieldtheory}, showed also that this basis provides the best control over prior volume effects when constraining $\Lambda$CDM parameters. 
For the counterterms, \texttt{velocileptors} implements a reparameterization in which they are expressed in terms of fractional corrections $\tilde{\alpha}_i$ to the linear theory. These corrections can be interpreted as multiplicative deviations from the baseline linear contribution and are transformed into the standard $\alpha_i$ coefficients internally. This approach allows setting transparent priors such that their contribution remains within physically motivated bounds.
For the stochastic terms, a common choice is to use Gaussian priors inspired by halo model arguments. The expected sizes of the stochastic amplitudes \(SN_0\), \(SN_2\), and \(SN_4\) can be approximated from satellite fractions \(f_\mathrm{sat}\) and virial velocity dispersions \(\sigma_v\)~\cite{maus2024analysisparametercompressionfullmodeling}, and these priors help mitigate parameter volume effects when the data alone are not informative enough. For these reasons, we ultimately adopted the physical basis as the baseline for our analysis. In the full-shape part of the analysis, we also chose to include only the monopole and quadrupole measurements; this implies fixing \(b_3\), \(\alpha_4\), \(\alpha_6\) and \(SN_4\) parameters to zero. 
The response of the new parameter \(\gamma_b\) to the choice of bias parameterization is non-trivial. Indeed, in our new approach, the effect of \(\gamma_b\) on the shape of the power spectrum leads to a degeneracy between the new parameter and the EFT nuisances since they are highly degenerate with amplitude parameters. As the data are already not well constraining the EFT parameters, we can expect even greater projection effects in both the physical and standard bases. To mitigate these effects, we opted for Halo Occupation Distribution (HOD) informed priors on the EFT set of parameters, as presented in~\cite{zhang2025hodinformedprioreftbasedfullshape, zhang2025enhancingdesidr1fullshape}. 

Another key aspect relevant for our work in the EFT context is the treatment of infrared (IR) resummation. It is well known that standard 1-loop perturbation theory fails to accurately predict both the position and amplitude of the BAO peak, particularly in real space. This limitation originates from the perturbative expansion itself. Several techniques have therefore been developed to mitigate this effect and properly resum the BAO signal. Our estimate of \(\gamma_b\) is largely driven by the amplitude of the BAO peak, making the impact of IR resummation potentially important for our analysis. We therefore repeated the analysis under different IR-resummation schemes, varying the exponential damping parameter \(\Sigma_\mathrm{BAO}^2\) around its fiducial value and modifying the filter used to separate the power spectrum into its wiggled and non-wiggled components. Further details are provided in the companion paper, and we verify that our results are robust to these choices, yielding consistent constraints across all tested configurations.

With the clustering signal, we simultaneously constrain the $\Lambda$CDM parameters that shape the linear power spectrum, namely $h$, $\ln(10^{10}A_s)$, $\omega_\mathrm{CDM}$, and $n_s$, applying broad flat priors. The baryon density is only weakly constrained by the power spectrum, so we fix it to $\omega_b = 0.02237$, consistent with the Planck 2018 value~\cite{Planck:2020} adopted in our fiducial cosmology. This choice does not affect our results. The total neutrino mass is fixed to the minimum allowed value $\sum m_\nu = 0.06~\mathrm{eV}$, consistent with the fiducial cosmology. However, in this analysis we are not interested in extracting cosmological information from the broadband shape of the power spectrum, and we marginalize over these parameters.

To implement the new density-based method for measuring \(H_0\), introduced in Section~\ref{sec:intro}, we must first isolate the physical processes to be used. In addition to the baryon fraction, we use the measurement of \(\omega_b\) from BBN and the measurement of \(\Omega_m\) from low-redshift Alcock–Paczyński parameters. Since these are the measurements employed in our density-based method, and are independent of the values inferred from the broadband information of the power spectrum, we denote them as \(\omega_b^{\mathrm{dens}}\) and \(\Omega_m^{\mathrm{dens}}\). If we further denote the dimensionless Hubble constant obtained from this method as \(h^{\mathrm{dens}}\), we can reparameterize the entire pipeline in terms of these three parameters simply through the following equation:
\begin{equation}
    \gamma_b = \frac{\omega_b^{\rm dens}}{{\Omega_m^{\rm dens} (h^{\rm dens})^2} - \omega_{\mathrm{massive}-\nu}}\,.
\end{equation}
$\Omega_m^{\rm dens}$ is constrained through the compressed-BAO component of the analysis, via the Alcock--Paczynski (AP) parameters $\alpha_\perp$ and $\alpha_\parallel$. While the AP parameters are sensitive to both the total matter density and the combination of $H_0$ with the sound horizon scale, we marginalize over the latter to eliminate contamination from sound horizon information.
Finally, we impose a Gaussian BBN constraint on $\omega_b^{\rm dens}$, with a width of 0.00055 matching recent BBN measurements \cite{schöneberg20242024bbnbaryonabundance} and flat priors on $h^{\rm dens}$ and $\Omega_m^{\rm dens}$. In Table~\ref{tab:cosmo_growth_params_priors} we summarize the priors used for the broadband cosmological parameters and for the new density-based parameters.
\begin{table}[ht]
\centering
\renewcommand{\arraystretch}{1.2}
\begin{tabular}{|c|c|}
\hline
 & \textbf{Priors} \\
\hline
$h$ & $\mathcal{U}[0.1, 1.0]$ \\
$\ln{(10^{10}A_s)}$ & $\mathcal{U}[2.0, 4.0]$ \\
$\omega_\mathrm{CDM}$ & $\mathcal{U}[0.08, 0.16]$ \\
$n_s$ & $\mathcal{U}[0.8, 1.2]$ \\
$\omega_b^{\mathrm{dens}}$ & $\mathcal{N}[\mu_{\omega_b}, 0.00055]$ \\
$\Omega_m^{\mathrm{dens}}$ & $\mathcal{U}[0.2, 0.5]$ \\

$h^{\mathrm{dens}}$& $\mathcal{U}[0.1, 1.0]$ \\
\hline
\end{tabular}
\caption{\justifying
Set of priors employed in the baseline analysis for the cosmological parameters and our new density-based method parameters. Uniform and Gaussian priors are denoted by $\mathcal{U}[\mathrm{min}, \mathrm{max}]$ and $\mathcal{N}(\mu, \sigma)$, respectively. The limits for uniform priors correspond to the limits of the training space of the emulator employed in the analysis (for emulator's details see Section~\ref{sec:HOD_pipeline}). The center
of the $\omega_b^{\textrm{dens}}$ prior, $\mu_{\omega_b}$, is chosen to be the true value of $\omega_b$ for the test under consideration, $\omega_b  = 0.02237$ for Abacus and the true $\omega_b$ value for the noiseless mock tests.}
\label{tab:cosmo_growth_params_priors}
\end{table}

\subsection{HOD informed priors}
\label{sec:HOD_pipeline}
To mitigate the strong degeneracies and projection effects associated with poorly constrained nuisance parameters in EFT modeling, we adopt the HOD-informed prior (HIP) approach developed by Zhang et al. in \cite{zhang2025enhancingdesidr1fullshape}. This method uses physically motivated priors on nuisance parameters derived from simulations of galaxy formation based on HOD models. By calibrating the EFT parameters against mock galaxy catalogs generated from HOD prescriptions applied to a suite of cosmological N-body simulations, the resulting priors reflect realistic galaxy–halo connections and significantly reduce the volume of unphysical parameter space.
To achieve this, Zhang et al. generated real-space mock galaxy catalogs for each DESI tracer (BGS, LRG, ELG, QSO) using different HOD models within a set of 32 AbacusSummit simulation boxes~\cite{garrison2016improving, garrison2018abacus, garrison2019high, garrison2021abacus, bose2022constructing}, which span a 7-dimensional cosmological parameter space. After converting these catalogs into redshift space and computing the galaxy power spectrum monopole and quadrupole for each realization, they fit the multipoles with the EFT model (specifically, the one-loop Eulerian perturbation theory implementation in \texttt{velocileptors}). During the fitting process, the cosmological parameters were kept fixed and so, only the nuisance parameters were fitted. They used these best-fit values to train a normalizing flow model, which learns a highly expressive joint prior distribution over the EFT nuisance parameters. These parameters are expressed in the \textit{physical} basis.
These priors were shown to successfully suppress projection effects, particularly in extended cosmological models (such as \(w_0w_a\)CDM), yielding constraints that are consistent with physical expectations and that recover the true cosmology from mock datasets. 

\begin{figure}[!ht]
    \centering
    \includegraphics[scale = 0.55]{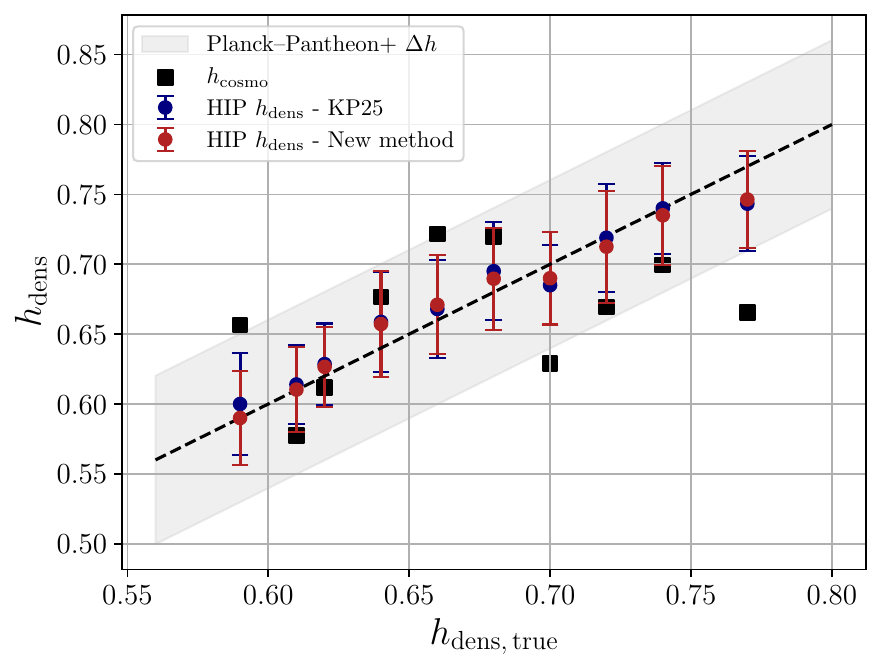}
    \caption{\justifying Validation test of the HOD-informed priors in scenarios where the growth and cosmological baryon fractions differ. The circles show the recovered values of \(h_\mathrm{dens}\) against their true inputs, \(h_\mathrm{dens, true}\), for both the KP25 splitting (blue) and the new method (red). The black squares show the value of the cosmological Hubble parameter \(h\) on the y-axis, as a function of the true \(h_\mathrm{dens}\) value. Their purpose is to illustrate the difference between the \(h_\mathrm{dens}\) value used and the underlying cosmological \(h\), showing that the pipeline successfully recovers the correct \(h_\mathrm{dens}\) with an error smaller than the difference from the cosmological value. The small residual effect of the priors, which slightly push density-based values toward their cosmological counterparts, is negligible compared to the statistical uncertainty (only 10-15\% of the error bar). The shaded gray band highlights the tension in \(H_0\) between Planck 2018 and Pantheon+.}
    \label{fig:hod_test_h}
\end{figure}

However, it is important to note that the simulations used to calibrate these priors were generated under cosmologies in which the growth baryon fraction matches the cosmological one. As such, it is essential to verify that the application of these priors does not introduce an artificial bias in the inference of \(\gamma_b\) towards the cosmological baryon fraction, especially when this parameter is treated as independent and allowed to vary.
To assess this potential issue, we performed an additional dedicated validation test using noiseless data vectors generated from cosmologies in which the growth and cosmological baryon fractions differ (Fig.~\ref{fig:hod_test_h}).

The choice to adopt HOD-informed priors comes with a significant performance drawback. Since these priors are not standard distributions but are instead represented as Normalizing Flows, it is not possible to analytically marginalize over the counterterms and stochastic terms. This requires sampling a fairly high-dimensional parameter space (50 parameters in the case of 6 redshift bins). To optimize the analysis, we therefore implemented the entire pipeline --- from theory modeling to likelihood definition and sampling ---in \texttt{Julia}~\cite{bezanson2015juliafreshapproachnumerical}, which allowed us to leverage high-performance, state-of-the-art tools. Making use of the same set of best-fit parameters provided by Zhang et al., we retrained the normalizing flows directly in \texttt{Julia} using the deep learning library \texttt{Lux.jl}~\cite{pal2023lux}, and interfaced them with the probabilistic programming language \texttt{Turing.jl}~\cite{turing1, turing2} to express our inference model in a compact and efficient way. The use of \texttt{Julia} also enabled us to take advantage of \texttt{Effort.jl}~\cite{bonici2025effortfastdifferentiableemulator}, an emulator capable of replicating any PT-based theory code and dramatically accelerating the full analysis. We used it to emulate the EFT model in the Eulerian formulation of \texttt{velocileptors}, specifically as implemented in the REPT class of the official DESI library, \texttt{desilike}\footnote{\url{https://desilike.readthedocs.io/en/latest/}}. Additionally, using the \texttt{AbstractCosmologicalEmulators.jl}~\cite{bonici2025effortfastdifferentiableemulator} and \texttt{Emulatorstrainer.jl}~\cite{bonici2025effortfastdifferentiableemulator} frameworks, the core of \texttt{Effort.jl} emulator, we trained an emulator for \(\sigma_8(z)\), since the growth of structures in our model now depends also on the new growth parameters, and an accurate estimate of \(\sigma_8\) is required to convert EFT parameters from the physical to the standard basis. All these tools support fast and accurate automatic differentiation of every component, allowing us to sample chains with any gradient-based sampler and to obtain best fit solutions using the efficient L-BFGS minimization algorithm.

\section{Validation tests} 
\label{sec:validation-tests}

We conducted two main types of validation tests. Having established the presence of projection effects, we first verified that we could adopt the same HOD-informed priors as in~\cite{zhang2025enhancingdesidr1fullshape} without introducing biases in the results. Subsequently, using the final pipeline, we performed a comparison between the previous KP25 splitting scheme and the new method introduced in this work, employing noiseless data vectors generated from a set of \(\Lambda\)CDM (AbacusSummit cosmologies~\cite{Maksimova21, garrison2021abacus}) and EDE cosmologies, covering a sufficiently wide range of \(h\) values. For both tests, in order to best assess the power of this method at the precision level of currently available data, we employed the DESI DR1 covariance matrices, performing the analysis on the six official DESI redshift bins (three bins for the Luminous Red Galaxies, one for the Emission Line Galaxies, one for the Quasars and the Bright Galaxy Sample)~\cite{Adame_2025, desicollaboration2025desi2024vfullshape, Forero_S_nchez_2025, Rashkovetskyi_2025}. We also restricted the full-shape part of the analysis to the same \(k\) range used in DESI Y1, \(0.02 \leq k \leq 0.2\;h\,\mathrm{Mpc}^{-1}\).
DESI is a redshift survey designed to map the three-dimensional distribution of matter in the Universe with unprecedented precision~\cite{desicollaboration2016desiexperimentiiinstrument, DESI_Collaboration_2022, miller2023opticalcorrectordarkenergy, 2024AJ....168..245P}. Operating on the 4-m Mayall Telescope at Kitt Peak National Observatory, DESI uses 5000 robotically controlled optical fibers to obtain spectra of millions of galaxies and quasars across $14{,}000\,\mathrm{deg}^2$ of sky~\cite{Guy_2023, schlafly2024surveyoperationsdarkenergy}. Its primary goal is to provide stringent constraints on the expansion history of the Universe and the growth of structure. The first cosmological results, including BAO measurements from the initial year of observations, were presented in 2024~\cite{desicollaboration2025datarelease1dark, Adame_2025}, and in 2025 the BAO results and cosmological parameter constraints from the second data release were published~\cite{desicollaboration2025desidr2resultsii, Anonymous_2025, andrade2025validationdesidr2measurements, casas2025validationdesidr2lyalpha}.

We adopted the NUTS sampler~\cite{hoffman2011nouturnsampleradaptivelysetting} and, to ensure efficient performance and good convergence (Gelman–Rubin criterion of \(1-R<0.01\)), we reparameterized the parameter space and the datavectors such that the variance along each dimension was normalized to unity. This allowed us to run shorter chains while still achieving a convergence of \(1-R<0.005\) and an average Effective Sample Size (ESS) of at least 1000 for the cosmological parameters, and \(1-R<0.01\) with an average ESS of at least 800 for the EFT parameters.

\begin{figure*}[!htpb]
    \centering
    \begin{subfigure}{0.45\textwidth}
        \includegraphics[width=\linewidth]{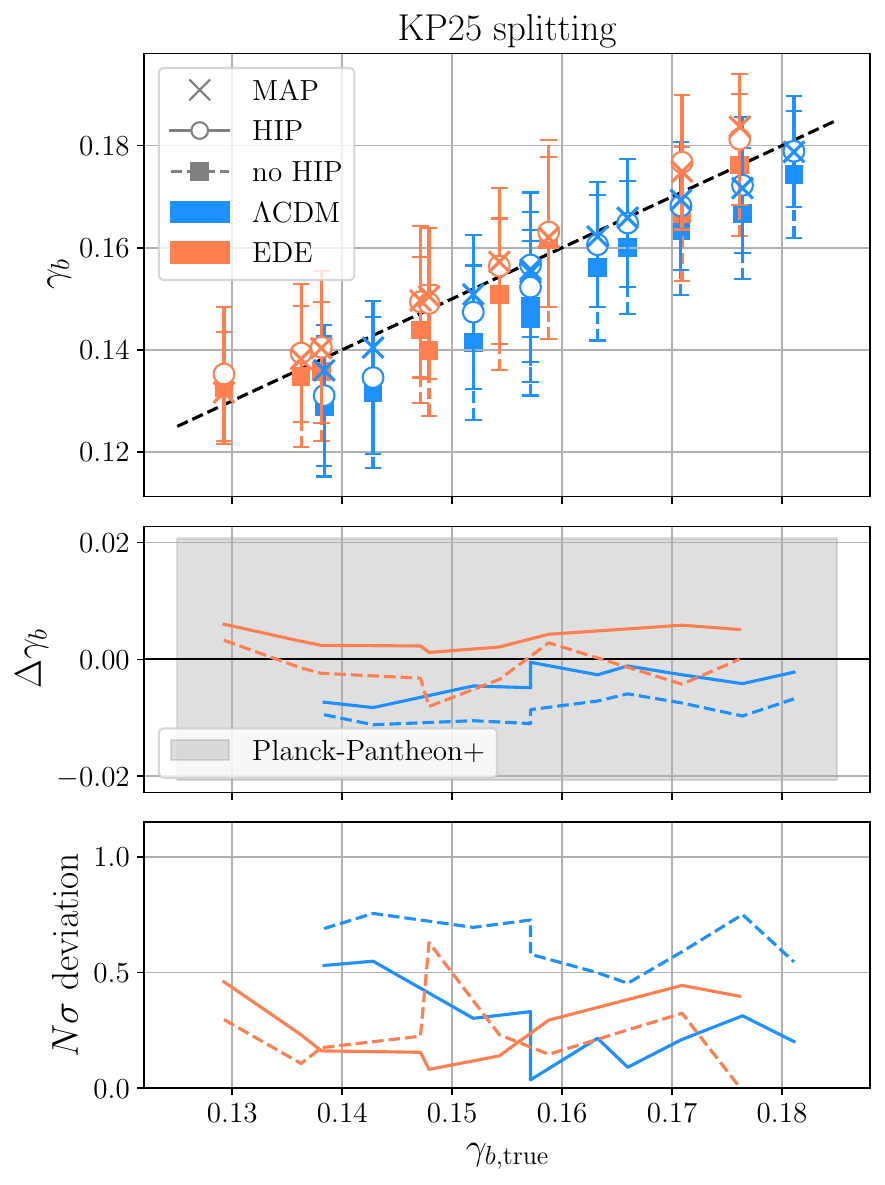}
    \end{subfigure}
    \begin{subfigure}{0.45\textwidth}
        \includegraphics[width=\linewidth]{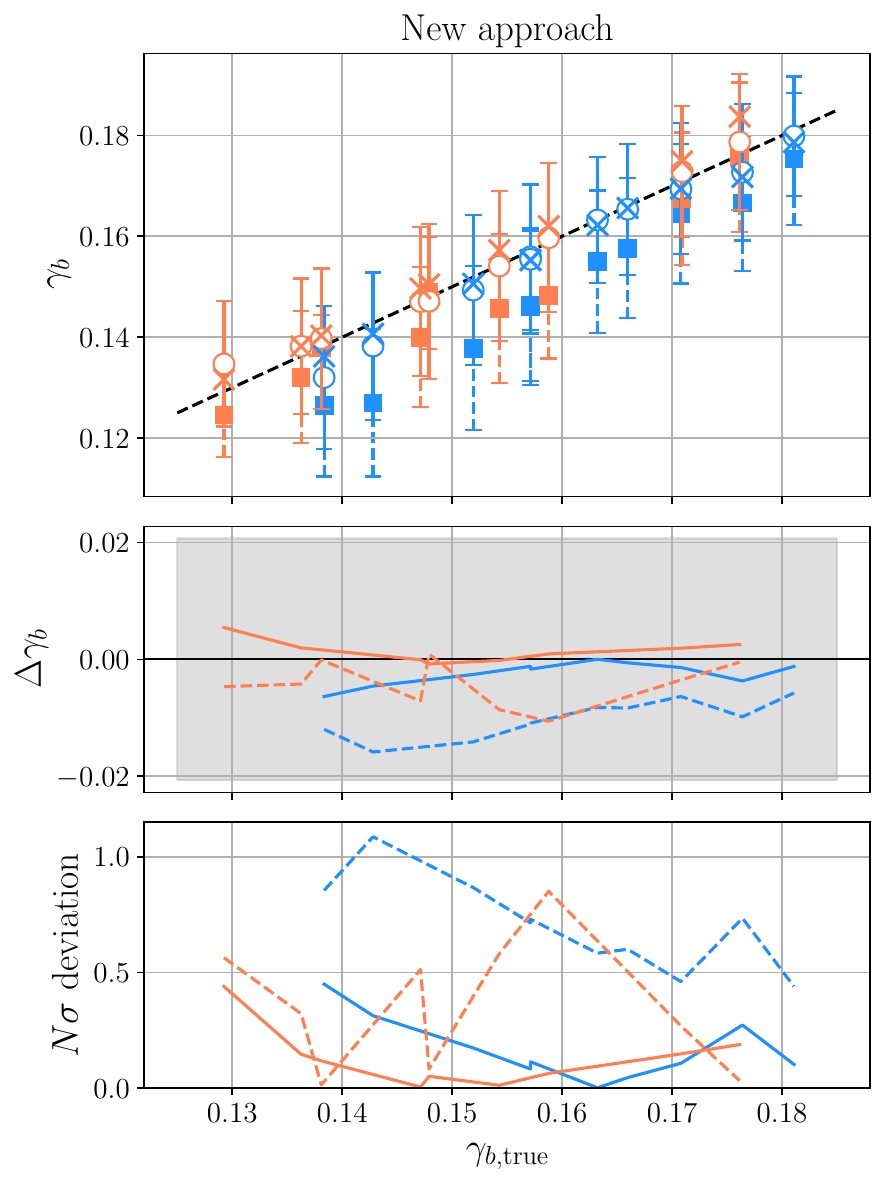}
    \end{subfigure}
    \begin{subfigure}{0.45\textwidth}
        \includegraphics[width=\linewidth]{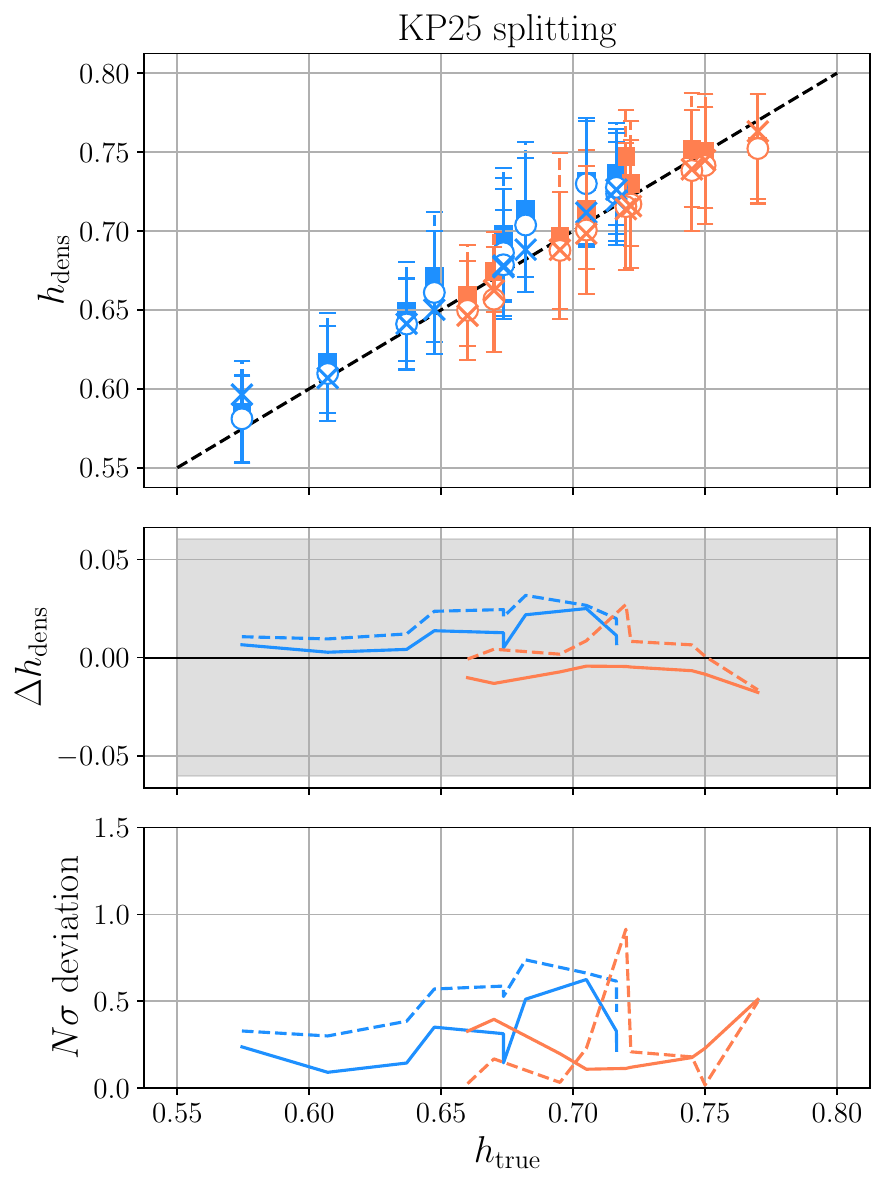}
    \end{subfigure}
    \begin{subfigure}{0.45\textwidth}
        \includegraphics[width=\linewidth]{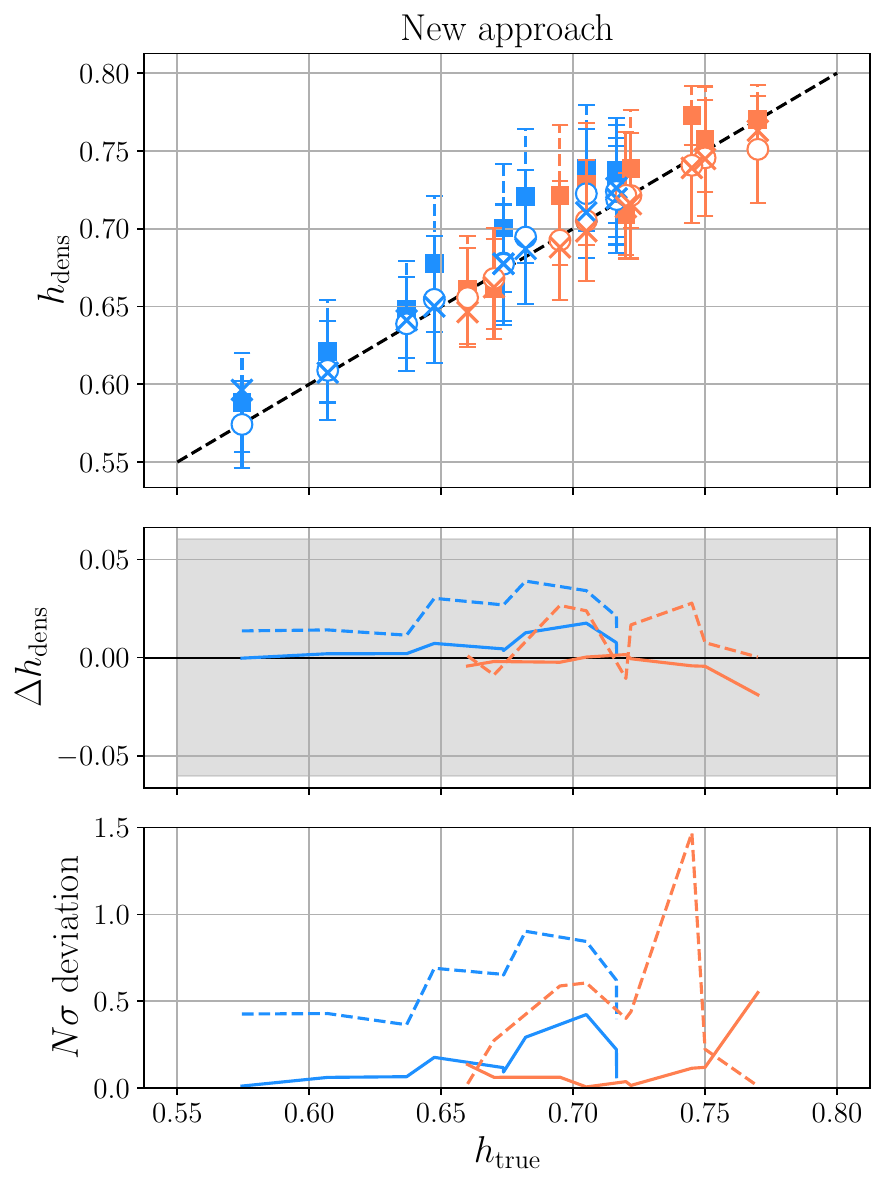}
    \end{subfigure}
    \vspace{-2mm}
    \caption{\justifying Comparison between the true and inferred \(\gamma_b\) and \(h_\mathrm{dens}\) from noiseless data vectors, for both \(\Lambda\)CDM (blue) and EDE (orange) cosmologies. Circles represent the marginalized constraints using the HOD-informed prior, while squares represent the constraints using the default DESI priors. Crosses show the maximum of the posterior. Results are shown for the KP25 splitting method (left) and the new perturbation-evolution approach (right). The bottom panels show the residuals \(\Delta\) with respect to the true value, with the shaded band marking the Planck–Pantheon+ tension, and the absolute deviation as number of \(\sigma\).}
    \label{fig:old_vs_mod_noiseless_gamma_b_h}
\end{figure*}

\subsection{HIP validation}
We verified the validity of the HOD priors in scenarios where the growth and cosmological baryon fractions do not match. We selected a set of simulations generated from different cosmologies in which the cosmological and growth baryon fractions coincide. We then performed a fit of all parameters except for \(\gamma_b\), which was fixed to a value different from the cosmological baryon fraction. We fixed \(\gamma_b\) to values up to \(~27\%\) different from the cosmological baryon fraction, which translates into \(h^\mathrm{dens}\) values up to \(~16\%\) different from the cosmological \(h\).
The resulting fit provides a set of cosmological and EFT parameters giving approximately the same power in the multipoles. Using these parameters, we generated new noiseless simulations of a non-physical Universe in which the growth and cosmological baryon fractions no longer coincide. This is equivalent to fixing \(\sigma_8\) to a different value and adjusting the other parameters accordingly.
As shown in Fig.~\ref{fig:hod_test_h}, the pipeline is able to recover the true \(h^\mathrm{dens}\) used to generate the data, even when this value is, in some cases, significantly different from the cosmological one. 

It is also evident that there exists a correlation between the bias and the cosmological Hubble parameter: when the cosmological Hubble parameter \(h\) exceeds \(h^\mathrm{dens}\), the resulting bias is positive; conversely, when it is lower, the bias becomes negative. This convinced us that the HIPs have a minor effect on the \(\gamma_b\) recovered from our pipeline, slightly pushing the values towards their cosmological counterparts. This effect is very small and, relative to the error bars, is negligible, amounting to only about 10–15\% of the standard deviation. For clarity, we also overplot a shaded gray area representing the tension between the \(H_0\) values from Planck2018~\cite{Planck:2020} and Pantheon+~\cite{Brout_2022}. Similarly, we verified that both \(\gamma_b\) and \(\Omega_m^\mathrm{dens}\) are unbiased, obtaining the same correlation between the sign of the residual bias and the cosmological parameter, but again with the bias that remains within the same 10–15\% of the corresponding \(\sigma\).

\begin{table*}[!th]
\centering
\renewcommand{\arraystretch}{1.2}
\setlength{\tabcolsep}{6pt}
\begin{tabular}{|c c c c c c c c|}
\hline
Cosmo & \(\omega_b\) & \(\omega_{c}\) & \(h\) & \(A_s\) & \(n_s\) & \(\omega_{ncdm}\) & \(\sigma_{8}\) \\
\hline
000 & 0.02237 & 0.1200 & 0.6736 & \(2.0830\times10^{-9}\) & 0.9649 & 0.0006442 & 0.807952 \\
130 & 0.02237 & 0.1200 & 0.6736 & \(1.6140\times10^{-9}\) & 0.9649 & 0.0006442 & 0.711201 \\
131 & 0.02237 & 0.1086 & 0.7165 & \(2.3146\times10^{-9}\) & 0.9649 & 0.0006442 & 0.807866 \\
139 & 0.02416 & 0.1128 & 0.7164 & \(2.1681\times10^{-9}\) & 0.9732 & 0.0006442 & 0.793003 \\
159 & 0.02206 & 0.1261 & 0.5967 & \(1.7977\times10^{-9}\) & 0.9673 & 0.0006442 & 0.718521 \\
167 & 0.02248 & 0.1400 & 0.7050 & \(1.7321\times10^{-9}\) & 0.9715 & 0.0006442 & 0.768419 \\
168 & 0.02157 & 0.1084 & 0.6070 & \(2.5378\times10^{-9}\) & 0.9275 & 0.0006442 & 0.871407 \\
169 & 0.02189 & 0.1222 & 0.6474 & \(1.8067\times10^{-9}\) & 0.9898 & 0.0006442 & 0.777925 \\
172 & 0.02282 & 0.1032 & 0.6369 & \(2.1259\times10^{-9}\) & 0.9012 & 0.0006442 & 0.765650 \\
177 & 0.02239 & 0.1344 & 0.6820 & \(2.0641\times10^{-9}\) & 1.0002 & 0.0006442 & 0.806026 \\
181 & 0.02169 & 0.1112 & 0.5745 & \(2.0651\times10^{-9}\) & 0.9336 & 0.0006442 & 0.730036 \\
\hline
\end{tabular}
\caption{AbacusSummit cosmologies chosen to test recovery of \(\gamma_b\).}
\label{tab:abacus_cosmos}
\end{table*}

This confirms that the information encoded in the HOD priors does not induce spurious constraints or biases on these parameters, and can therefore be safely employed within our energy density framework.
\subsection{Noiseless data-vectors test}
We have conducted a series of validations on noiseless data vectors generated from a set of \(\Lambda\)CDM and EDE cosmologies. In this analysis, we applied both the previous KP25 splitting scheme and the new method introduced in this work, in order to carefully assess their differences in terms of the precision with which they recover the value of the Hubble constant, and the residual bias from projection effects. 
We selected the \(\Lambda\)CDM cosmologies from the AbacusSummit suite, specifically the 10 listed in Table~\ref{tab:abacus_cosmos}. For the EDE case, we fixed the fiducial cosmological parameters as studied in~\cite{Hill20} (\(h=0.7219\), \(\omega_b=0.02253\), \(\omega_c=0.1306\), \(A_s=2.215\times10^{-9}\), \(n_s=0.9889\), \(f_\mathrm{EDE}=0.122\), \(\log_{10}{z_c}=3.562\), \(\theta_{i,\mathrm{scf}}=2.83\)), and then considered nine variations of this cosmology by varying \(h\) while keeping \(\omega_b\) and \(\Omega_m\) fixed, so that changes in \(h\) would fully translate into variations of the baryon fraction. With the \(\Lambda\)CDM cosmologies we covered the \(h\) range 0.57–0.72, while the EDE cosmologies spanned the range 0.66–0.77. 
In Fig.~\ref{fig:old_vs_mod_noiseless_gamma_b_h}, we compare the true value of \(\gamma_b\) and \(h\) with the value inferred by our pipeline for the noiseless \(\Lambda\)CDM and EDE data vectors. We find that the two methods yield equally precise results: \(\sim8\%\) for \(\gamma_b\) and \(\sim5\%\) for \(h_\mathrm{dens}\). However, the KP25 method appears to retain a slightly larger bias compared to the new approach. 
The resulting \(\Delta h\) and \(\Delta \gamma_b\) values remain entirely negligible compared to the constraining power of current data for both methods, as well as to the most relevant \(H_0\) tension. We additionally plot the biases in \(\gamma_b\) and \(h\) in units of their error. When using the default prior on the biases and counterterms, we find significant projection effects leading to \(\sim 0.5 \sigma\) biases in the marginalized posteriors. The HOD-informed prior effectively mitigates these projection effects, reducing the biases on the marginalized
means back to the \(\lesssim 0.1\sigma\) range.
Thus, we have proven that the new approach is capable of inferring the correct value of \(\gamma_b\) --- and therefore \(H_0\) --- regardless of whether the Universe is \(\Lambda\)CDM or EDE.

The old splitting and our new method both extract their primary information from the BAO amplitude, but they differ in how the baryon fraction affects the smooth shape of the power spectrum. The two approaches are therefore complementary, and the fact that they yield consistent results at this level of precision provides a confirmation of the robustness of this approach.

\section{Conclusions}
\label{sec:conclusions}

In this paper we have developed a new framework to measure the baryon fraction from galaxy clustering, with the explicit goal of enabling energy density based determinations of the Hubble constant. Previous work introduced the idea of extracting the baryon fraction from the amplitude of the baryon signal in the galaxy power spectrum, but implemented it through an approximate template-based transfer-function splitting that neglected part of the baryon contribution to the growth of structure.

Here, we have proposed an improved, physically self-consistent method that embeds the growth baryon fraction parameter $\gamma_{b}$ directly into the linear perturbation evolution equations of \texttt{CAMB}. This ensures that both baryons and CDM respond coherently to variations in $\gamma_{b}$ across all cosmic epochs, while preserving the correct thermal history. We integrated this implementation into a complete EFT-based full-shape analysis pipeline, further enhanced with HOD-informed priors to mitigate degeneracies and projection effects. 

Through validation tests on noiseless data vectors from both $\Lambda$CDM and EDE cosmologies, we demonstrated that the new approach achieves comparable precision to the previous splitting scheme (about $8\%$ on $\gamma_{b}$ and $5\%$ on $h$), while substantially reducing systematic biases. Importantly, it maintains the ability to recover the correct value of $\gamma_{b}$ and thus $H_{0}$ even in extended cosmologies, confirming its robustness. Although the statistical precision is not improved relative to the previous method, the new approach offers a cleaner and more accurate physical description and therefore represents a more reliable foundation for future applications.
Overall, our results show that incorporating $\gamma_{b}$ at the perturbation level provides a robust and self-consistent route to baryon-fraction measurements. Combined with BBN constraints on $\omega_{b}$ and Alcock–Paczynski determinations of $\Omega_{m}$, this establishes a viable way to measure $H_{0}$ independently of the sound horizon.

The application of this method already represents a significant innovation in the context of the \(H_0\) tension, with substantial prospects for improvement. Since the precision on \(H_0\) is entirely determined by the precision on \(\gamma_b\), which improves with increasing survey volume, this methodology will have the opportunity to demonstrate its strength in ongoing and future deep surveys such as DESI, Euclid~\cite{2025, Euclid20}, Rubin Observatory’s Legacy Survey of Space and Time (LSST)~\cite{Ivezi__2019}, and Roman~\cite{2020JATIS...6d6001M}. DESI will cover a comoving volume of $\sim 60\,h^{-3}\,\mathrm{Gpc}^{3}$ \cite{DESISV}, nearly ten times that of BOSS, reducing the statistical uncertainty on $\gamma_{b}$ significantly and enabling $H_{0}$ determinations precise enough to clearly separate the SH0ES and CMB/LSS values at $>3\sigma$. Comparable sensitivity is expected from Euclid, with a volume of $\sim 55\,h^{-3}\,\mathrm{Gpc}^{3}$ \cite{Euclid20}, while additional progress will come from all-sky spectroscopic surveys such as SPHEREx \cite{Spherex14} and the Roman Space Telescope \cite{Eifler21}. Meanwhile, BAO or full-shape-based analyses tied to $r_\mathrm{d}$ are forecast to reach sub-percent precision, with uncertainties below $\sigma_{H_{0}} < 0.5\,\mathrm{km\,s^{-1}\,Mpc^{-1}}$ \cite{Farren22, Ivanov23b}. The combination of these complementary methods will provide stringent cross-checks, offering a decisive opportunity to test the robustness of $\Lambda$CDM and to explore possible new physics behind the $H_{0}$ discrepancy.

\section*{Data Availability}
The data used in this work are public as part of DESI Data Release 1 (details at \url{https://data.desi.lbl.gov/doc/releases/}). The data points corresponding to the figures
are available on Zenodo at
\url{https://zenodo.org/uploads/17686137}. Code is available on github \url{https://github.com/drew2799/fbCAMB}.

\acknowledgments

We thank Antony Lewis for useful conversations about the structure of \texttt{CAMB}.
AK was supported as a CITA National Fellow by the Natural Sciences and Engineering Research Council of Canada (NSERC), funding reference \#DIS-2022-568580.
WP acknowledges support from the Natural Sciences and Engineering Research Council of Canada (NSERC), [funding reference number RGPIN-2025-03931] and from the Canadian Space Agency.
Research at Perimeter Institute is supported in part by the Government of Canada through the Department of Innovation, Science and Economic Development Canada and by the Province of Ontario through the Ministry of Colleges and Universities.
This research was enabled in part by support provided by Compute Ontario (computeontario.ca) and the Digital Research Alliance of Canada (alliancecan.ca).
This material is based upon work supported by the U.S. Department of Energy (DOE), Office of Science, Office of High-Energy Physics, under Contract No. DE–AC02–05CH11231, and by the National Energy Research Scientific Computing Center, a DOE Office of Science User Facility under the same contract. Additional support for DESI was provided by the U.S. National Science Foundation (NSF), Division of Astronomical Sciences under Contract No. AST-0950945 to the NSF’s National Optical-Infrared Astronomy Research Laboratory; the Science and Technology Facilities Council of the United Kingdom; the Gordon and Betty Moore Foundation; the Heising-Simons Foundation; the French Alternative Energies and Atomic Energy Commission (CEA); the National Council of Humanities, Science and Technology of Mexico (CONAHCYT); the Ministry of Science, Innovation and Universities of Spain (MICIU/AEI/10.13039/501100011033), and by the DESI Member Institutions: \url{https://www.desi.lbl.gov/collaborating-institutions}. Any opinions, findings, and conclusions or recommendations expressed in this material are those of the author(s) and do not necessarily reflect the views of the U. S. National Science Foundation, the U. S. Department of Energy, or any of the listed funding agencies.
The authors are honored to be permitted to conduct scientific research on I'oligam Du'ag (Kitt Peak), a mountain with particular significance to the Tohono O’odham Nation.





\appendix
\section{Author Affiliations}
\label{sec:affiliations}
\textsuperscript{4}Lawrence Berkeley National Laboratory, 1 Cyclotron Road, Berkeley, CA 94720, USA\\ 
\textsuperscript{5}Department of Physics, Boston University, 590 Commonwealth Avenue, Boston, MA 02215 USA\\ 
\textsuperscript{6}Dipartimento di Fisica ``Aldo Pontremoli'', Universit\`a degli Studi di Milano, Via Celoria 16, I-20133 Milano, Italy\\ 
\textsuperscript{7}INAF-Osservatorio Astronomico di Brera, Via Brera 28, 20122 Milano, Italy\\ 
\textsuperscript{8}Department of Physics \& Astronomy, University College London, Gower Street, London, WC1E 6BT, UK\\ 
\textsuperscript{9}Instituto de F\'{\i}sica, Universidad Nacional Aut\'{o}noma de M\'{e}xico,  Circuito de la Investigaci\'{o}n Cient\'{\i}fica, Ciudad Universitaria, Cd. de M\'{e}xico  C.~P.~04510,  M\'{e}xico\\ 
\textsuperscript{10}University of California, Berkeley, 110 Sproul Hall \#5800 Berkeley, CA 94720, USA\\ 
\textsuperscript{11}Institut de F\'{i}sica d’Altes Energies (IFAE), The Barcelona Institute of Science and Technology, Edifici Cn, Campus UAB, 08193, Bellaterra (Barcelona), Spain\\ 
\textsuperscript{12}Departamento de F\'isica, Universidad de los Andes, Cra. 1 No. 18A-10, Edificio Ip, CP 111711, Bogot\'a, Colombia\\ 
\textsuperscript{13}Observatorio Astron\'omico, Universidad de los Andes, Cra. 1 No. 18A-10, Edificio H, CP 111711 Bogot\'a, Colombia\\ 
\textsuperscript{14}Institut d'Estudis Espacials de Catalunya (IEEC), c/ Esteve Terradas 1, Edifici RDIT, Campus PMT-UPC, 08860 Castelldefels, Spain\\ 
\textsuperscript{15}Institute of Cosmology and Gravitation, University of Portsmouth, Dennis Sciama Building, Portsmouth, PO1 3FX, UK\\ 
\textsuperscript{16}Institute of Space Sciences, ICE-CSIC, Campus UAB, Carrer de Can Magrans s/n, 08913 Bellaterra, Barcelona, Spain\\ 
\textsuperscript{17}Fermi National Accelerator Laboratory, PO Box 500, Batavia, IL 60510, USA\\ 
\textsuperscript{18}Institut d'Astrophysique de Paris. 98 bis boulevard Arago. 75014 Paris, France\\ 
\textsuperscript{19}IRFU, CEA, Universit\'{e} Paris-Saclay, F-91191 Gif-sur-Yvette, France\\ 
\textsuperscript{20}Department of Physics, University of Michigan, 450 Church Street, Ann Arbor, MI 48109, USA\\ 
\textsuperscript{21}University of Michigan, 500 S. State Street, Ann Arbor, MI 48109, USA\\ 
\textsuperscript{22}Department of Physics, The University of Texas at Dallas, 800 W. Campbell Rd., Richardson, TX 75080, USA\\ 
\textsuperscript{23}NSF NOIRLab, 950 N. Cherry Ave., Tucson, AZ 85719, USA\\ 
\textsuperscript{24}Department of Physics and Astronomy, University of California, Irvine, 92697, USA\\ 
\textsuperscript{25}The Ohio State University, Columbus, 43210 OH, USA\\ 
\textsuperscript{26}Sorbonne Universit\'{e}, CNRS/IN2P3, Laboratoire de Physique Nucl\'{e}aire et de Hautes Energies (LPNHE), FR-75005 Paris, France\\ 
\textsuperscript{27}Departament de F\'{i}sica, Serra H\'{u}nter, Universitat Aut\`{o}noma de Barcelona, 08193 Bellaterra (Barcelona), Spain\\ 
\textsuperscript{28}Center for Cosmology and AstroParticle Physics, The Ohio State University, 191 West Woodruff Avenue, Columbus, OH 43210, USA\\ 
\textsuperscript{29}Department of Astronomy, The Ohio State University, 4055 McPherson Laboratory, 140 W 18th Avenue, Columbus, OH 43210, USA\\ 
\textsuperscript{30}Instituci\'{o} Catalana de Recerca i Estudis Avan\c{c}ats, Passeig de Llu\'{\i}s Companys, 23, 08010 Barcelona, Spain\\ 
\textsuperscript{31}Space Sciences Laboratory, University of California, Berkeley, 7 Gauss Way, Berkeley, CA  94720, USA\\ 
\textsuperscript{32}Instituto de Astrof\'{i}sica de Andaluc\'{i}a (CSIC), Glorieta de la Astronom\'{i}a, s/n, E-18008 Granada, Spain\\ 
\textsuperscript{33}Departament de F\'isica, EEBE, Universitat Polit\`ecnica de Catalunya, c/Eduard Maristany 10, 08930 Barcelona, Spain\\ 
\textsuperscript{34}Department of Physics and Astronomy, Sejong University, 209 Neungdong-ro, Gwangjin-gu, Seoul 05006, Republic of Korea\\ 
\textsuperscript{35}Abastumani Astrophysical Observatory, Tbilisi, GE-0179, Georgia\\ 
\textsuperscript{36}Department of Physics, Kansas State University, 116 Cardwell Hall, Manhattan, KS 66506, USA\\ 
\textsuperscript{37}Faculty of Natural Sciences and Medicine, Ilia State University, 0194 Tbilisi, Georgia\\ 
\textsuperscript{38}CIEMAT, Avenida Complutense 40, E-28040 Madrid, Spain\\ 
\textsuperscript{39}Department of Physics \& Astronomy, Ohio University, 139 University Terrace, Athens, OH 45701, USA\\ 
\textsuperscript{40}National Astronomical Observatories, Chinese Academy of Sciences, A20 Datun Road, Chaoyang District, Beijing, 100101, P.~R.~China\\

\bibliographystyle{JHEP}
\bibliography{References}



\end{document}